\documentclass[english,preprint]{revtex4}
\usepackage[T1]{fontenc}
\usepackage[latin1]{inputenc}
\usepackage{amsmath}
\usepackage{graphicx}

\makeatletter

\usepackage{babel}
\makeatother
\begin{document}

\title{Correlation of worldwide markets' entropies: time-scale approach}

\author{José A.O. Matos$^{1,2,3}$, Sílvio M.A. Gama$^{1,3}$, Heather J.
Ruskin$^{4}$, Adel Sharkasi$^{4}$, Martin Crane$^{4}$}

\address{$^{1}$ Centro de Matemática da Universidade do Porto, Edifício dos
Departamentos de Matemática da FCUP, Rua do Campo Alegre 687, 4169-007
Porto, Portugal}

\address{$^{2}$ Grupo de Matemática e Informática, Faculdade de Economia
da Universidade do Porto, Rua Roberto Frias, 4200-464 Porto, Portugal}

\address{$^{3}$ Departamento de Matemática Aplicada, Faculdade de Ciências
da Universidade do Porto, Rua do Campo Alegre 687, 4169-007 Porto,
Portugal}

\address{$^{4}$ School of Computing, Dublin City University , Dublin 9, Ireland}



\begin{abstract}
We use a new method of studying the Hurst exponent with time and scale
dependency. This new approach allow us to recover the major events
affecting worldwide markets (such as the September 11th terrorist
attack) and analyze the way those effects propagate through the different
scales. The time-scale dependence of the referred measures demonstrates
the relevance of entropy measures in distinguishing the several characteristics
of market indices: \char`\"{}effects\char`\"{} include early awareness,
patterns of evolution as well as comparative behaviour distinctions
in emergent/established markets.
\end{abstract}
\maketitle

\section{Introduction}

\subsection{Goals}

The goal of this study is the analysis of stock exchange world indices
searching for signs of coherence and/or synchronization across the
set of studied markets.

We have expanded the scope of previous work on the PSI-20 (Portuguese
Standard Index), since results there \cite{Matos2004} seemed to provide
a basis for a wider ranging study of coherence and entropy.

With that purpose we applied econophysics techniques related to measures
of {}``disorder''/complexity (entropy) and a newly proposed \cite{Matos2006}
generalization of Detrended Fluctuation Analysis. As a measure of
coherence among a selected set of markets we have studied the eigenvalues
of the correlation matrices for two different set of markets, exploring
the dichotomy represented by emerging and mature markets.

The data used in this study was taken daily for a set of worldwide
market indices. As it is usual in this kind of analysis \cite{Mantegna2000}
we base our results on the study of log returns $\eta_{i}=\log\frac{x_{i}}{x_{i-1}},$
where $\eta_{i}$ is the log return at time step $i$.

\section{Entropy}

The Shannon entropy for blocks of size $m$ for an alphabet of $k$
symbols is \cite{Shannon1948} \begin{equation}
\overset{\sim}{H}(m)=-\sum_{j=0}^{k^{m}-1}p_{j}\log p_{j},\label{eq:entropy-sequence}\end{equation}
the entropy of the source is then\begin{equation}
\overset{\sim}{h}=\lim_{m\rightarrow\infty}\frac{\overset{\sim}{H}(m)}{m}.\label{eq:entropy-source}\end{equation}

This definition is attractive for several reasons: it is easy to calculate
and it is well defined for a source of symbol strings. In the particular
case of returns, if we choose a symmetrical partition we know that
half of the symbols represent losses and half of the symbols represent
gains. If the sequence is predictable, we have the same losses and
gains sequences repeated everytime, the entropy will be lower; if
however all sequences are equally probable the uncertainty will be
higher and so it will be the entropy. Entropy is thus a good measure
of uncertainty.

This particular method has problems, the entropy depends on the choice
of encoding and it is not a unique characteristic for the underlying
continuous time series. Also since the number of possible states grows
exponentially with $m$, after a short number of sequences in practical
terms it will become difficult to find a sequence that repeats itself.
This entropy is not invariant under smooth coordinate changes, both
in time and encoding. This is a strong handicap for its adoption into
financial time series study.

We have applied the Shannon entropy for blocks of size $5$ and an
alphabet of $50$ symbols, to a set of markets previously studied.
We should recall that using blocks of size $5$ corresponds to a week
in trading time. Notice also that we have only considered trading
days, like what we do in all other analysis, so we ignore any holidays
or days where the market was closed.

It should be noted that results are robust to the choice of the total
number of bins (the size of our alphabet). That is, we have repeated
the analysis with a different choice of the number of partitions yielding
similar results.

In order to enhance the time dependence of results we have evaluated
the entropy of the set for periods of $100$ trading days (roughly
corresponding to half a year). The motivation for this analysis is
to study the time evolution of entropy.%
\begin{figure}
\includegraphics[width=0.4\columnwidth,bb = 0 0 200 100, draft, type=eps]{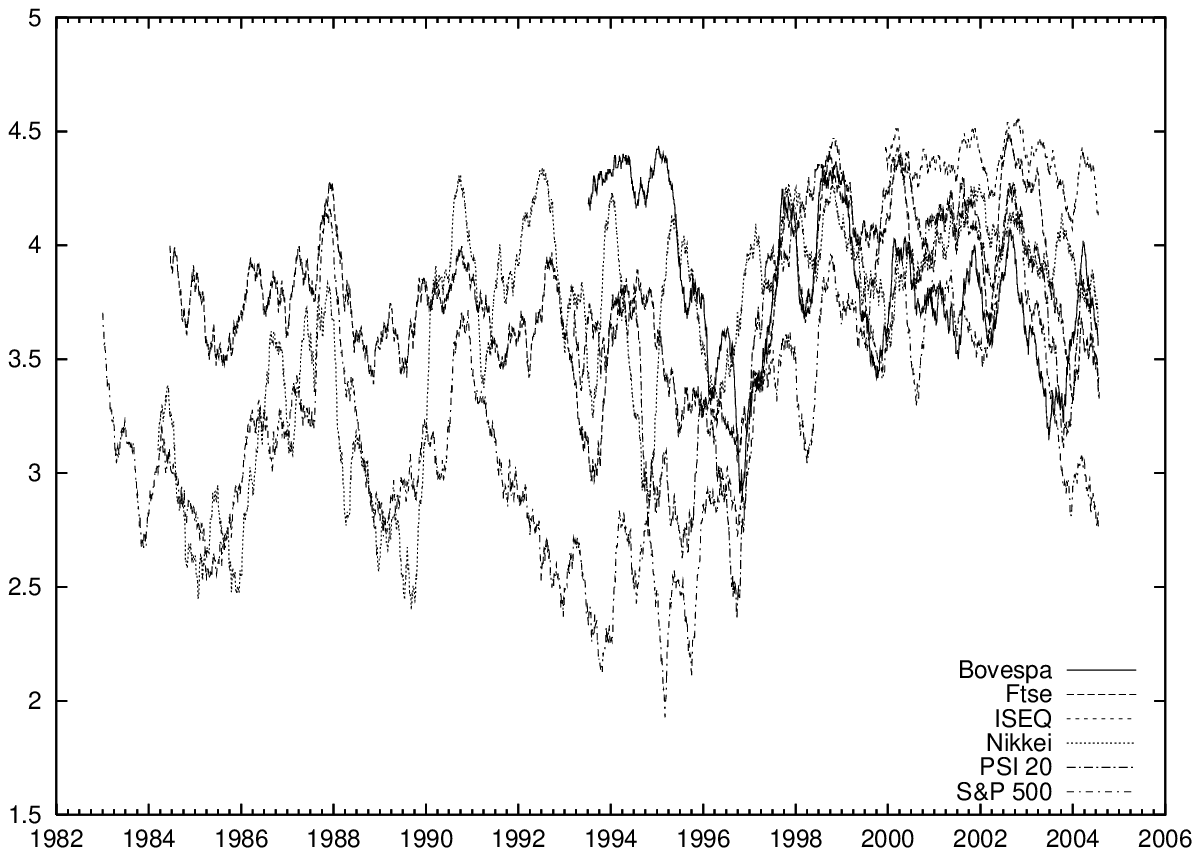}

\caption{\label{cap:Weekly-entropy}Weekly entropy for various market indexes.}
\end{figure}

The results displayed in Figure~\ref{cap:Weekly-entropy} show improved
coherence (i.e. reduced entropy) after $1997$ as compared with previous
periods for all markets. Higher entropy implies less predictability,
in general, although the nature of shocks qualifies this statement
to some extent. The notable feature of this graphic is that both mature
and developing markets are affected similarly which suggests that
global behaviour patterns are becoming more coherent or linked because
of the progressive globalisation of markets. This is in line with
the findings of \cite{Matos2006} where we found the Hurst exponent
for different markets to be decreasing with time.

\section{Time and Scale Hurst exponent}

\subsection{Method characterisation}

The general idea behind this method is the study of the Hurst exponent
as a function of both time and scale. In practical terms this method
is a simple expansion of the {}``windowed'' DFA applied in \cite{Matos2004}.
Instead of fixing $s$ we let it be a variable. The Hurst exponent,
$H(t,s)$, for time $t$ and scale $s$, is evaluated as the Hurst
exponent obtained using the DFA \cite{Peng1994}, for the interval
$[t-s/2;t+s/2]$.

Implications are wider than for a simple DFA. The general idea is
to essentially invert the process and take $H(s,t)$ as the focus
of the analysis with the DFA being an implementation detail. The other
candidate to evaluate the Hurst exponent in the sub-intervals is the
wavelet \cite{Percival2000}. In both cases $H$ is recovered as a
power of the scale, inside each sub-interval.

Recalling the most important equation in DFA we have the detrended
fluctuation function as:

\[
F(t)\sim t^{H},\]
 where $H$ is the Hurst exponent.

From the above condition we know that $s/2+1\leq t\leq T-s/2$, where
$T$ is the time series length. In what follows the maximum scale
we consider is $s=T/4$ as for large scales we essentially recover
the Hurst exponent for the whole series.

A major concern in this work was to guarantee that exponents obtained
through DFA were meaningful. For that reason we have used the same
procedure as in \cite{Matos2004}, we have controlled the quality
of the fits assuring that the regression coefficients of the linear
least squares fits were near unity for all studied markets. If we
would not do this, the results would be unreliable, since the underlying
time series is not well described by a fractional Brownian motion.
To this combination of the DFA with time and scale dependency, we
apply the term TSDFA (Time and Scale DFA).

\subsection{Examples}

Here we study some examples of the technique applied to several international
markets. We choose these because they display details that are either
unique or shared with other markets and contribute to understand the
differences and similarities that TSDFA emphasises.

Traditionally we distinguish between developed and emergent markets,
the distinction varies depending on the source and of the applied
criteria. A more in depth discussion of this issue is found in Section~\ref{sec:TSDFA-Results}.

\subsubsection{Nikkei}

As an illustration of the method we worked with Nikkei $225$ data
ranging from 1990 to 2005. Nikkei was chosen because it is a well
known and studied financial index.

\begin{figure}
\includegraphics[width=0.4\columnwidth,bb = 0 0 200 100, draft, type=eps]{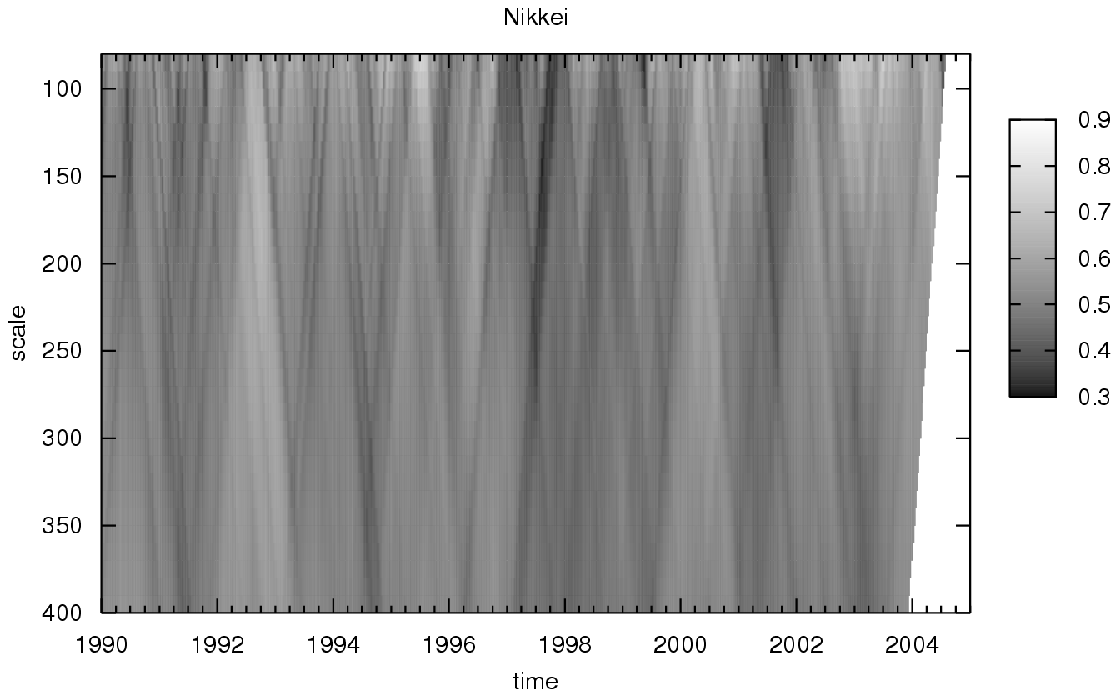}\hfill{}\includegraphics[width=0.4\columnwidth,bb = 0 0 200 100, draft, type=eps]{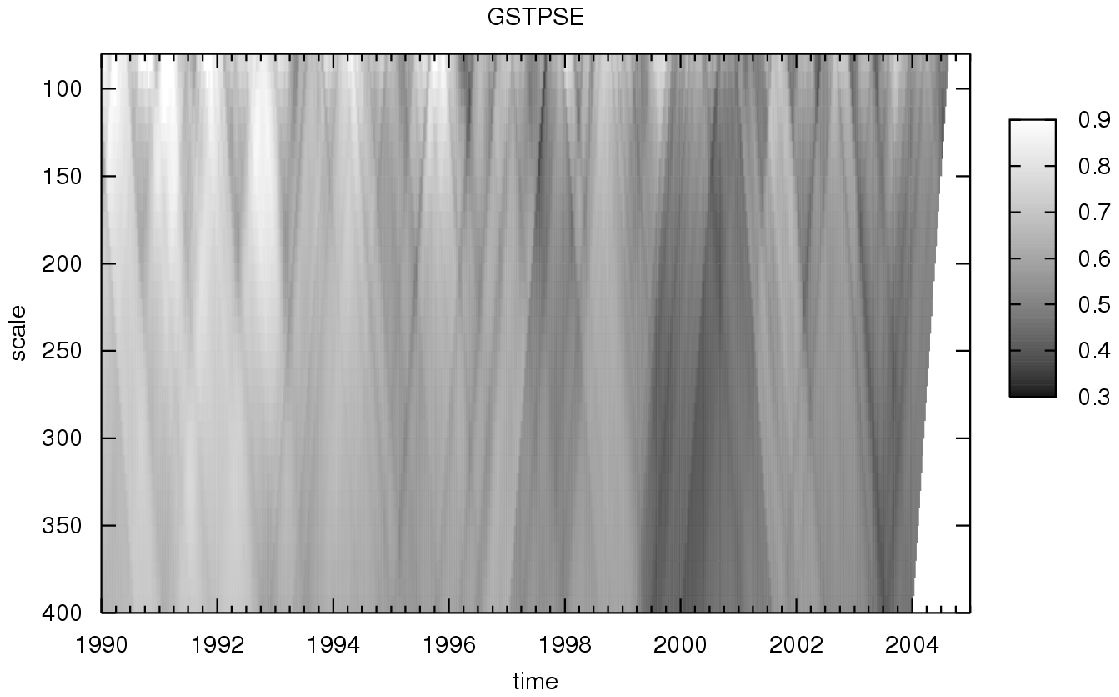}

\includegraphics[width=0.4\columnwidth,bb = 0 0 200 100, draft, type=eps]{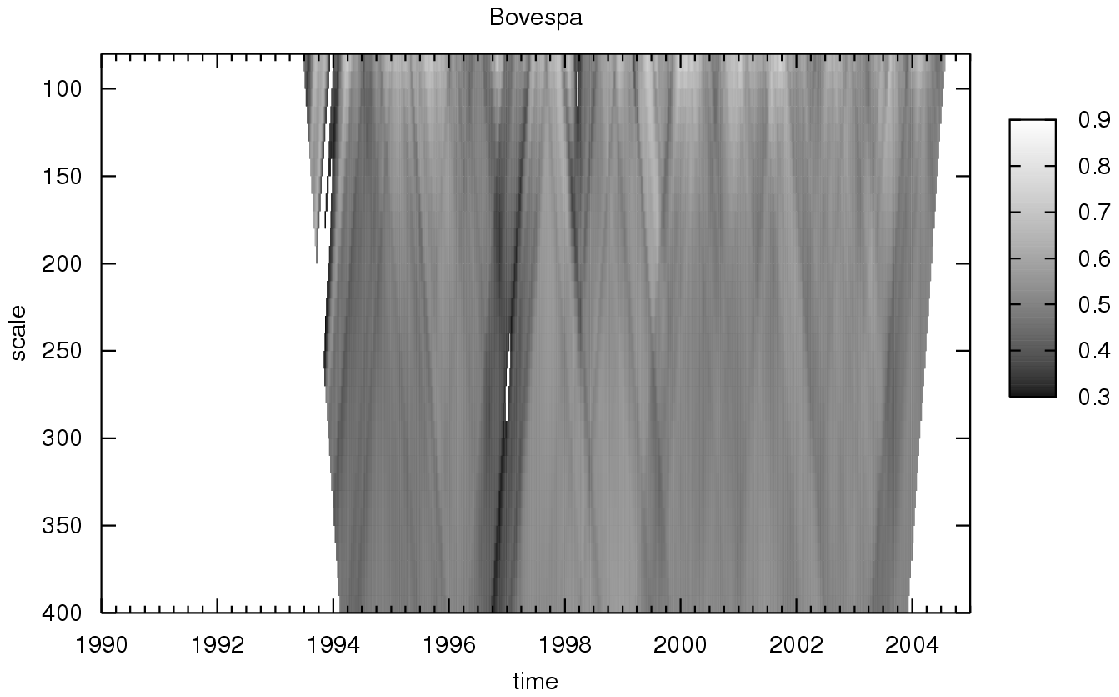}\hfill{}\includegraphics[width=0.4\columnwidth,bb = 0 0 200 100, draft, type=eps]{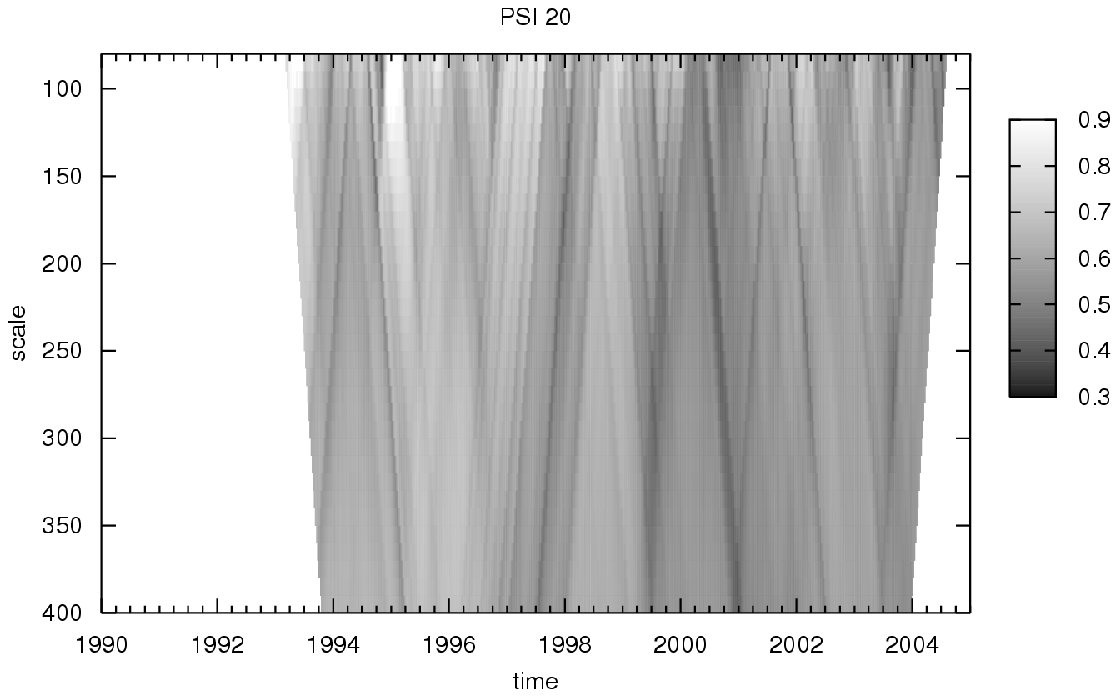}

\caption{\label{cap:TSDFA}TSDFA applied to several markets. The scale (in
trading days) is represented by the y-axis; the time is represented
in x-axis (years).}
\end{figure}

The graph resulting from application of the TSDFA method is shown
in Figure~\ref{cap:TSDFA}.%
The graphic represents as a contour plot, with exponents in range
$[0.3;0.9]$, the series studied from $1990-2005$ and the scale between
$100$ and $400$ trading days. In this work we adopted these fixed
ranges since this representation permits ready comparison with other
indices calculated.

In the Nikkei graphic (Figure~\ref{cap:TSDFA}) we can see that persistence
is exhibited with the index normally around $0.5.$ This reflects
a healthy borderline, of values near $0.5$, and is to be expected
since Nikkei is a mature market. In recent years we see a stripe that
crosses all scales in year $2000$, at the same time as the DotCom
crash.

We have another stripe that starts in the fourth quarter of $2001$
but does not go through all scales. Another period of high values
of $H$ starts for short scales in the third quarter of $2002$, after
a global crash and reaches large scales in $2004$.

\subsubsection{GSTPSE (Canada)}

As seen in Figure~\ref{cap:TSDFA}, the market shows two distinct
periods, before and after $1997$. Before $1997$ we see high values
of Hurst exponent over all scales. After that time, all the regions
of high Hurst exponents are bounded in time and the background turns
out to be what we expect from a mature market, with the Hurst exponent
around $0.5$.

There are two stripes, for high values of $H$, after $1997$ that
cross all scales, one in $1998$ and another starting around September
$2001$ and travelling forward for higher scales in time.

\subsubsection{Bovespa (Brazil)}

Bovespa, the São Paulo Stock Exchange Index, is known for its high
volatility and is generally considered an emergent market. In Figure~\ref{cap:TSDFA}
we see an erratic behaviour with $H$ either near or above $0.5$
and the corresponding stripes crossing together, back and forward
in time, at all scales. There are two stripes, for high values of
$H$, that start from short scales respectively in $1997$ (Asian
crashes) and $1998$ (global crash) which merge for large scales.
There is another, for $H>0.5$, stripe that walks through all scales
and starts for short scales around September $2001$.

\subsubsection{PSI-20 (Portugal)}

Unmodified DFA, the predecessor of TSDFA, was applied to PSI-20 in
\cite{Matos2004}. In Figure~\ref{cap:TSDFA} we see the results
of applying TSDFA to this market, from establishment of series in
$1993$.

Initial stages are both antipersistent and subject to extreme values
of the Hurst exponent. We can identify two stripes with a stable (higher)
value of the Hurst exponent, during $1998$, and another walking forward
in time starting, for short scales, next to September $2001$. Notice
that this stripe is so strong that it overlaps other stripes forming
in the neighbourhood.

The overall strength of the TSDFA is to provide further conclusions
over those drawn earlier concerning the market progression to mature
behaviour and its responses times, clearly different from the initial
position.

\subsection{Features}

As can be seen in Figure~\ref{cap:TSDFA} there are several notable
features of the plots produced by TSDFA:

\begin{itemize}
\item We can distinguish mature markets by the persistence and stability
of $H$ values around $0.5$, most of the time.
\item We can distinguish emergent markets by the persistence and stability
of $H$ values above $0.5$.
\item For some periods, a phase transition appears to occur, sometimes observable
across all scales, sometimes across partial scales only. This is reflected
in the spikes which either point to lower or to large scales;
\item \emph{A priori}, we expected smooth variations of $H$ for large scales
since we are taking into account more data values and therefore we
expect greater robustness to sudden changes of the data. This was
already observed in the results obtained for PSI-20 and is confirmed
by all the examples.
\item Markets evolve in time, the Canadian case is a notable example of
this, where we observe a shift from emergent to mature features. Although
not so dramatic for all other cases we see over time a decrease in
the values of the Hurst exponent.
\item There are events that change the Hurst exponent behaviour that can
be seen in most/all markets. The September 11th 2001 is the most striking
case that can be seen in all Figures.
\item Clearly, the behaviour is dependent both on time and scale, indicative
of the multifractal background, so that details obtained are richer
than those obtained by calculation of the Hurst exponent directly.
This is to be expected since the Hurst exponent is a summary measure,
or index, of the data and this is the observed behaviour for financial
markets (see \cite{Lux2004}).
\end{itemize}

\section{\label{sec:TSDFA-Results}Results}

\subsection{Classification of global markets}

The classification of markets into mature or emergent is not a simple
issue. The International Finance Corporation (IFC) uses income per
capita and market capitalisation relative to GNP for classifying equitity
markets. If either 1) a market resides in a low or middle-income economy,
or 2) the ratio of the investable market capitalisation to GNP is
low, then the IFC classifies the market as emerging, otherwise the
classification is mature.

It seems clear from the results, obtained from TSDFA, that we can
distinguish different markets classes. The difference in behaviour
is visible with the application of TSDFA. The most active, and mature,
markets show a persistence of behaviour near $H=0.5$ while the newer,
emergent, markets show a persistence of higher values of $H$. The
diversity of behaviours does not stop here,  there are markets which
show an hybrid behaviour between these two states.

The classification that we propose has thus three states:

\begin{description}
\item [{(clearly)~mature}] these market have a persistence of $H$ around
$0.5$. The presence of regions with higher values of $H$ is limited
to small periods and is well defined both in time and scale.
\item [{(clearly)~emergent}] these market have a persistence of $H$ well
above $0.5$. The presence of regions with values of $H$ around $0.5$
is well defined both in time and scale.
\item [{hybrid}] unlike the two previous case the distinction between the
mature and emergent phases is not well determined, with the behaviour
seemingly mixing at all scales.
\end{description}

\subsection{Data}

We have considered, in this study, the major and most active markets
worldwide from America (North and South), Asia, Africa, Europe and
Oceania. All the data on the respective market indices is public and
came from Yahoo Finance (\url{finance.yahoo.com}). We have considered
the daily closure as the value for the day, to obviate any time zone
difficulties.

The choice of the markets used in this study was driven by the goal
of studying major markets across the world in an effort to ensure
that tests and conclusions could be as general as possible. Hence
from the results we have divided the markets according to mature:
AEX General (Netherlands); Dow Jones (U.S.); CAC 40 (France); FTSE
100 (United Kingdom); DAX (Germany); S\&P 500 Index (U.S.); Nasdaq
(U.S.); Seoul Composite (South Korea); Nikkei 225 (Japan); NYSE Composite
Index (United States) and Stockholm General (Sweden). The list of
hybrid markets is smaller: All Ordinaries (Australia); Bovespa (Brazil);
S\&P TSX Composite (Canada); NZSE 10 (New Zealand); Madrid General
(Spain) and Swiss Market (Switzerland).

All the other markets from our study behave as emergent: ATX (Austria);
BEL-20 (Belgium); BSE 30 (India); CMA (Egypt); All Share (Sri Lanka);
Hang Seng (Hong Kong); IPSA (Chile); ISEC Small Cap (Ireland); ISEC
Small Cap Techno (Ireland); Irish SE Index (Ireland); Jakarta Composite
(Indonesia); KFX (Denmark); KLSE Composite (Malaysia); Karachi 100
(Pakistan); MerVal (Argentina); MIBTel (Italy); IPC (Mexico); OSE
All Share (Norway); PSE Composite (Philippines); PSI 20 (Portugal);
PX50 (Czech Republic); Shanghai Composite (China); Straits Times (Singapore);
TA-100 (Israel); Taiwan Weighted (Taiwan) and ISE National-100 (Turkey).

\section{Conclusions}

We applied the TSDFA (Time and Scale Detrended Fluctuation Analysis)
to study each market evolution in time and sometimes, as seen in some
markets, we observe a switch from developed to mature state. TSDFA
is used to compare the results of sets of markets and to establish
classes that display similar behaviour at any given time. This classification
allow us to distinguish events that affect several markets from other
local occurrences that only affect a single market as well as some
events that are reflected worldwide (the Asian tigers crashes, the
9/11 already cited above, the Madrid bomb attack in 2004, among others).
The resulting classification is in agreement with another based on
wavelet analysis proposed in \cite{SRC+06}.

One of the interesting outcomes is that, in spite of the results showing
differences between known emergent markets and established ones, we
found convergence of entropy behaviour in recent years among the worldwide
markets studied. We have found that more and more markets exhibit
a more mature behaviour. A plausible explanation for this phenomenon
is the progressive globalization of financial markets.

\bibliographystyle{plain}
\bibliography{Matos}

\begin{thebibliography}{1}

\bibitem{Lux2004}
T.~Lux.
\newblock Detecting {M}ulti-{F}ractal {P}roperties in {A}sset {R}eturns: {A}n
  {A}ssessment of the '{S}caling {E}stimator'.
\newblock {\em International Journal of Modern Physics}, 15:481 -- 491, 2004.

\bibitem{Mantegna2000}
R.N. Mantegna and H.E. Stanley.
\newblock {\em An {I}ntroduction to {E}conophysics}.
\newblock Cambridge University Press, Cambridge, 2000.

\bibitem{Matos2004}
J.A.O. Matos, S.M.A. Gama, H.J. Ruskin, and J.A.M.S. Duarte.
\newblock An econophysics approach to the {P}ortuguese {S}tock
  {I}ndex--{PSI}-20.
\newblock {\em Physica A}, 342(3-4):665--676, 2004.

\bibitem{Matos2006}
J.A.O. Matos, S.M.A. Gama, A.~Sharkasi, H.J. Ruskin, and M.~Crane.
\newblock Temporal and {S}cale {DFA} {A}pplied to {S}tock {M}arkets.
\newblock In preparation, 2006.

\bibitem{Peng1994}
C.-K. Peng, S.V. Buldyrev, S.~Havlin, M.~Simons, H.E. Stanley, and A.L.
  Golderberger.
\newblock On the mosaic organization of {DNA} sequences.
\newblock {\em Phys. Rev. E}, 49:1685--1689, 1994.

\bibitem{Percival2000}
D.~Percival and A.~Walden.
\newblock {\em Wavelet methods for time series analysis}.
\newblock Cambridge University Press, 2000.

\bibitem{Shannon1948}
C.~Shannon.
\newblock A mathematical theory of communication.
\newblock {\em Bell System Technical Journal}, 27:379--423, 1948.

\bibitem{SRC+06}
A.~Sharkasi, H.J. Ruskin, M.~Crane, J.A.O. Matos, and S.M.A. Gama.
\newblock A wavelet-based method to measure stages of stock market development.
\newblock In preparation, 2006.

\end{thebibliography}

\end{document}